\documentclass[reprint,prx,superscriptaddress]{revtex4}

\usepackage[english]{babel}
\usepackage{graphicx}
\usepackage{dcolumn}
\usepackage[T1]{fontenc} 
\usepackage{amsmath}
\usepackage{bm}

\begin{document}
\title{Probing Permanent Dipoles in CdSe Nanoplatelets \\ with Transient Electric Birefringence}

\author{Ivan Dozov}
\email{ivan.dozov@u-psud.fr}
\affiliation{Laboratoire de Physique des Solides, Universit\'e Paris-Saclay, CNRS, Universit\'e Paris-Sud, UMR 8502, 91405 Orsay, France.}
\author{Claire Goldmann}
\affiliation{Laboratoire de Physique des Solides, Universit\'e Paris-Saclay, CNRS, Universit\'e Paris-Sud, UMR 8502, 91405 Orsay, France.}
\author{Patrick Davidson}
\affiliation{Laboratoire de Physique des Solides, Universit\'e Paris-Saclay, CNRS, Universit\'e Paris-Sud, UMR 8502, 91405 Orsay, France.}
\author{Benjamin Ab\'ecassis}
\email{benjamin.abecassis@ens-lyon.fr}
\affiliation{Laboratoire de Physique des Solides, Universit\'e Paris-Saclay, CNRS, Universit\'e Paris-Sud, UMR 8502, 91405 Orsay, France.}
\affiliation{Univ Lyon, ENS de Lyon, CNRS UMR 5182, Université Claude Bernard Lyon 1, Laboratoire de Chimie, F69342, Lyon, France}

\date{\today}

\begin{abstract}
Zinc-blende CdSe semiconducting nanoplatelets (NPL) show outstanding quantum confinement properties thanks to their small, atomically-controlled, thickness. For example, they display extremely sharp absorption peaks and ultra-fast recombination rates that make them very interesting objects for optoelectronic applications. However, the presence of a ground-state electric dipole for these nanoparticles has not yet been investigated. We therefore used transient electric birefringence (TEB) to probe the electric dipole of 5-monolayer thick zinc-blende CdSe NPL with a parallelepipedic shape. We studied a dilute dispersion of isolated NPL coated with branched ligands and we measured, as a function of time, the birefringence induced by DC and AC field pulses. The electro-optic behavior proves the presence of a large dipolar moment (> 245 D) oriented along the length of the platelets. We then induced the slow face-to-face stacking of the NPL by adding oleic acid. In these stacks, the in-plane dipole components of consecutive NPL cancel whereas their normal components add. Moreover, interestingly, the excess polarizability tensor of the NPL stacks gives rise to an electro-optic contribution opposite to that of the electric dipole. By monitoring the TEB signal of the slowly-growing stacks over up to a year, we extracted the evolution of their average length with time and we showed that their electro-optic response can be explained by the presence of a 80 D dipolar component parallel to their normal. In spite of the $\bar{4}$3m space group of bulk  zinc-blende CdSe, these NPL thus bear an important ground-state dipole whose magnitude per unit volume is twice that found for wurtzite CdSe nanorods. We discuss the possible origin of this electric dipole, its consequences for the optical properties of these nanoparticles, and how it could explain their strong stacking propensity that severely hampers their colloidal stability.  
\end{abstract}
\maketitle

\section{Introduction}
Among colloidal nanocrystals, zinc-blende CdSe semiconducting nanoplatelets (NPL) have recently emerged as a new class of particles with ground-breaking optical properties. \cite{Ithurria2008,Ithurria2011,Nasilowski2016,lhuillier2015} These nanoparticles are the two-dimensional equivalent of quantum dots but quantum confinement occurs only along the dimension parallel to their atomically controlled thickness. As a consequence, they display original optical features such as extremely sharp absorption peaks, large binding energies, and fast recombination rates. \cite{Ithurria2011} These properties could be useful in a variety of optoelectronic applications such as LED. \cite{chen2014,Giovanella2018} Moreover, two key features of this class of materials have recently been discovered. Ultra-fast Forster Resonance Energy Transfer (FRET) occurs at the nanoscale between NPL \cite{rowland2015,Guzelturk2015}. FRET is particularly fast in the case of nanoplatelets and even faster than Auger recombination \cite{kunneman2013} which plagues quantum dots performances in devices \cite{klimov2000,Guzelturk2016}. Furthermore, NPL display strongly anisotropic, directed photo-luminescence while their absorption is isotropic. \cite{Scott2017,Gao2017a} These properties can potentially overcome current limitations observed in devices through engineering exciton flow in assemblies of NPL. However, exploiting these new (directional) optical properties requires organizing NPL orderly in a three-dimensional fashion, both in solution and on a surface. Several different methods have been used to induce macroscopic alignment of NPL such as stretching unidirectionally a NPL-polymer elastic composite \cite{Beaudoin2015,Beaudoin2017,Cunningham2016} or the use of evaporation-mediated self-assembly on a liquid \cite{Gao2017a} or solid substrate \cite{Ma2018a}. Another way to orient particles is by applying an electric field. This strategy can be very efficient if the particles bear an electric dipole since the dipoles will orient parallel to the field. CdSe nanorods have been shown to orient over large distances in electric fields \cite{Ryan2006,Carbone2007} but such a strategy has never been applied for NPL.

The presence of a ground state electric dipole in CdSe NPL is unknown. Such a dipole can arise due to the difference of electro-negativity between Cd and Se. Depending on the crystallographic structure and the shape of the particles, this modulated electronic distribution can yield a permanent dipole if the centers of mass of the negative and positive ionic lattices do not coincide. For wurtzite-type CdSe nanocrystals, theoretical calculations have shown that a permanent dipole should exist along the c-axis \cite{Rabani2001}  and electro-optical experiments have measured important dipolar moments in CdSe wurtzite quantum dots \cite{Blanton1997} and nanorods. \cite{Li2003} This type of crystalline structure is intrinsically polar but important dipolar moments have also been reported for zinc-blende type nano-structures whose cubic symmetry theoretically forbids the emergence of a significant dipole. \cite{Shim1999} The values of dipoles for CdSe nanocrystals range from 41 to 98 D for particles with diameters between 2.7 and 5.6 nm and increase with size. \cite{Blanton1997,Shim1999,Kortschot2014} For wurtzite, the measured dipole ranged between 25 and 50 D for spheres and can be as large as 210 D for nanorods. \cite{Li2003} All these values have been measured using dielectric spectroscopy or transient electric birefringence in the case of nanorods. Larger values, up to 500 D, have been extracted from cluster-size distribution measurements \cite{Klokkenburg2007} but these values strongly depend on the type of model used for the inter-particle potential. Though some reports link the dipolar moment with the total volume of the nanoparticles, there is no universally accepted scaling relation between the two quantities and recent simulations have shown that other factors such as ligand structure could have a more important effect. \cite{Greenwood2018}

The existence of a permanent dipole can impact both the optical properties and the self-assembly of nanocrystals. Ground-state dipoles can induce important effects on optical properties since the electric field within the crystal breaks the inversion symmetry, which has important consequences on the electronic states and hence on the emission spectrum. \cite{Schmidt1997} Differences in absorption spectra between one and two-photon excitations can notably be difficult to explain without evoking ground-state dipoles due to the mixing of states. Dipolar coupling between adjacent colloidal nanocrystals can also induce giant enhancement of the absorption cross-section in close-packed films compared to isolated nanoparticles. \cite{Geiregat2013} Dipole-dipole interactions dramatically impact colloidal self-assembly at the nanoscale. \cite{Talapin2007,Batista2015} For example, typical dipolar structures such as chains have already been observed in semi-conducting nanoparticle assemblies \cite{Dollefeld2001a,Klokkenburg2007,Shanbhag2007,Cho2005,Baskin2012}. 

It is thus important to determine whether or not CdSe NPL bear a permanent dipole. The crystallographic structure of CdSe NPL is zinc-blende. \cite{Ithurria2008,Chen2015} The top and bottom planes of the NPL are [001] basal planes with cadmium atoms linked to carboxylates ligands. \cite{Singh2018} These ligands provide the NPL with an initially moderate colloidal stability \cite{Jana2015} which can be improved with an “entropic” ligand. \cite{Yang2016} We have previously shown that CdSe NPL can assemble into a variety of structures depending on the assembly conditions such as giant micro-needles \cite{Abecassis2014} or (twisted) nano-ribbons \cite{Jana2016,Jana2017}. In these assemblies, NPL are stacked one on top of each other in a face to face fashion. \cite{Jana2015} Excitonic coupling in assemblies of NPL induces modifications to the fluorescence lifetime \cite{Guzelturk2014a} and causes the emergence of new peaks at low temperatures \cite{Tessier2013,Diroll2018}

Transient electric birefringence (TEB) is a technique of choice to investigate dipoles in colloidal nanoparticles. This method consists in submitting a colloidal suspension of the particles of interest to time-dependent electric fields and in recording its field-induced birefringence. \cite{Benoit1951} The TEB signal carries a specific signature when the particles have a permanent electric dipole.  
Indeed, since the pioneering work of Peterlin and Stuart \cite{Peterlin1939}, it was successfully used to answer similar questions in colloidal dispersions of various kinds of nanoparticles such as clays \cite{Shah1963,Holzheu2002,Dozov2011b,Arenas-Guerrero2016,Arenas-Guerrero2018}, mineral nanorods \cite{delacotte2015}, viruses \cite{Kramer1992}, protein fibrils \cite{Rogers2006}, macromolecules \cite{Frka-Petesic2014}, and mixtures of rod-like and spherical colloids \cite{Mantegazza2005}. Recently, this technique has also been applied to investigate the field-induced alignment of CdSe nanorods \cite{Mohammadimasoudi2016} or the orientation of transition metal dichalcogenide nanodiscs. \cite{Rossi2015}   

Here, we use TEB to measure ground-state dipoles in colloidal CdSe nanoplatelets. We first recall the theoretical background necessary to understand the experiments that were conducted. We then present our results on isolated CdSe NPL that are well dispersed in solution by the use of entropic ligands. The optical response of the dispersion to short square pulses of direct-current field and the relaxation dynamics when the field is switched off indicate that the most important contribution to the electric-field-induced orientation is due to a large permanent dipole in the plane of the NPL. This is confirmed by the low and high frequency alternating current experiments which are consistent with a permanent dipole larger than 245 D.  However, these experiments alone cannot give an estimate of the out-of-plane component of the dipole. To do so, we induced the slow stacking of the NPL by the addition of oleic acid. When the NPL stack, the TEB signal not only increases in intensity with time but also changes sign. This is due to the fact that, in NPL stacks, the in-plane dipolar components cancel out since adjacent NPL assemble with in-plane dipoles in opposite directions whereas out-of-plane components add. Furthermore, the geometry and the polarizability of the objects responding to the field change when the NPL stack, which gives rise to an unusual situation where the major components of the permanent and induced dipoles are orthogonal. 

\section{Theoretical background}
\subsection{Electric-field-induced birefringence of anisotropic objects}
When a strong electric field \( \bm{E} \) is applied to a dispersion of anisotropic particles, they are aligned by the field, on average, along or perpendicular to its direction, depending on the anisotropy of the electric properties of the particles. There are two different mechanisms of orientation: one is due to the polarizability anisotropy of the particle and the other is present only if the particle bears a permanent dipole. 


The total electric energy of a particle is then given by:
\begin{equation}
U^{p}(\Omega)=-\bm{\mu \cdot E} - \frac{1}{2}\bm{E \cdot \alpha \cdot E},
\label{eq:energy}
\end{equation}
where \( \bm{\alpha}\) is the (excess) polarizability tensor of the particles, \(\bm{\mu}\) is their permanent dipole moment, and \( \Omega\) denotes the set of Euler angles defining the particle orientation in the laboratory frame. 

When the dispersion is isotropic, as in our case, the particles reorient individually under the field action, with preferred average alignment defined by the sign of the particle electric anisotropy. Due to the revolution symmetry of the system around the field direction, \( \bm{e}\), the induced order is uniaxial. In the simplest case when the particles have cylindrical symmetry around some particle axis \( \bm{p}\), i.e. for rods and disks, the induced order is described by the scalar order parameter \( S(E)\) which depends on the field-induced and permanent dipoles of the particles (see supplementary information for the expression of \(S(E)\)).

At small particle volume fraction, \(\Phi \ll 1\), the induced nematic-like orientational order can be probed by measuring the equilibrium value of the field-induced birefringence of the colloidal dispersion \cite{OKonski1959}:
\begin{equation}
   \Delta n(E)=\Phi \Delta n^pS(E).
   \label{eq:biref}
 \end{equation}
 Here, \(\Delta n(E)=n_\parallel(E) - n_{\bot}(E) \) is the induced birefringence where \( \parallel\) and \(\bot\) denote the direction of light polarization with respect to \(\bm{e}\) and the refractive index \(n_i\) is related to the polarizability of the dispersion, \(\bm{\alpha^{opt}}\), in the optical frequency range by \(n_i(E) \propto \sqrt{\alpha_i^{opt}(E)}\). Similarly, the specific birefringence of the particle, defined as the birefringence of the dispersion for perfectly oriented particles (S = 1) and extrapolated to \(\Phi=1\), is \(\Delta n^p=n_\parallel^p - n_{\bot}^p\) where \( \parallel\) and \(\bot\) denote the direction of light polarization with respect to the symmetry axis \(\bm{p}\) of the particle. The specific birefringence is defined by both the internal structure of the particle (the “intrinsic” birefringence, which is only related to the optical anisotropy of the particle material) and its shape (the “form” birefringence, which also depends on the solvent refractive index).
For rods or disks in a weak field, the induced order is quadratic in the field \cite{OKonski1959}: 
\begin{equation}
  S(E)\simeq \frac{1}{15} \Delta A E^2 \quad \textrm{with} \quad \Delta A = \frac{1}{(kT)^2}\left(\mu_{\parallel}^2-\frac{1}{2}\mu_{\bot}^2\right)+\frac{1}{kT}\Delta \alpha, 
  \label{eq:order}
\end{equation}
where \(\Delta \alpha=\alpha_\parallel - \alpha_{\bot} \) is the anisotropy of the excess polarizability and the subscripts denote the components of \(\bm{\alpha}\) and \(\bm{\mu}\) parallel and perpendicular to \(\bm{p}\) (see figure 1.B.).


When the field-coupling coefficient \(\Delta A \) is positive, the particles align with their symmetry axis parallel to the field. For \( \Delta A < 0\), however, the induced order is negative, \(S<0\), and the symmetry axis of the particle tends to align perpendicular to the field. Thus, the sign of the induced birefringence gives important qualitative information about the strength and orientation of the permanent and induced dipole moments of the particle.

\subsection{Time-dependent electric field}
More quantitative information about the dipole moments can be retrieved from the time-response of the birefringence when the field varies in time, e.g. under AC voltage with variable frequency \cite{Thurston1969} or under pulsed DC field \cite{Tinoco1959}. Indeed, the different contributions to \( \Delta A\) have different relaxation behaviors because the rotational diffusion coefficient \(D^r\) depends on the orientation of the rotation axis, leading to parallel and perpendicular components, denoted \(D^r_{\parallel}\) and \(D^r_{\bot}\), respectively. Moreover, the transient behaviors of the dipole moment and the polarizability are qualitatively different because their couplings with the field are respectively linear and quadratic (see Eq.\ref{eq:energy}). Indeed, upon fast inversion of the sign of the field, the contribution of the polarizability to the energy does not change; the particle would then keep the same orientation. However, the energy of the permanent dipole changes its sign, resulting in a “head-to-tail” reorientation of the particle and therefore a transient change of the birefringence. \\

Here, we compare the rise and decay of the birefringence induced with square unipolar pulses \cite{Tinoco1959,OKonski1957}  and we study the frequency dependence of the DC and AC responses to bursts of sinusoidal voltage \cite{Peterlin1939,Thurston1969}. These techniques have been widely used for the study of the permanent and induced dipoles of colloidal particles with effective rotational symmetry, i.e. for disks and rods. But, the CdSe platelets lack this symmetry and a more general approach should be used for the interpretation of the TEB data \cite{Kalmykov2009}. However, to simplify the interpretation, we will approximate in the following the rotational diffusion of the platelet by considering that it behaves like a cylindrical particle, e.g. with revolution symmetry around one of its axes. We assume that the rotational diffusion coefficients around the two other axes, \(D^r_{\bot}\), are equal. Therefore, depending on the value of the diffusion coefficient for rotation around the effective symmetry axis, \(D^r_{\parallel}\), the platelet or the stack of platelets can be considered either as a rod with \(D^r_{\parallel} > D^r_{\bot}\) or as a disk with \(D^r_{\parallel} \leq D^r_{\bot}\), which greatly simplifies the data analysis.
The calculation of the diffusion coefficients (see SI for details) shows that the best uniaxial approximations, for both platelets and stacks, is that of a rod. For this geometry, the formulas describing the transient birefringence are well known.
For square pulses and weak fields (Kerr regime), i.e. for small enough induced order parameter, \( S(E) \ll 1\), when the field is switched on, the birefringence increases and reaches an equilibrium value given by: 
\begin{equation}
  \Delta n^{e}(E)= \Phi \Delta n^p S^e(E) = \frac{1}{15}\Phi \Delta n^p (p_{\parallel} - p_{\bot} + q) E^2=C_K^0 E^2
  \label{eq:deltane}
\end{equation}
with 
\begin{equation}
  C_K^0 = \lim_{E \to 0} \frac{\Delta n^e(E)}{E^2}; \quad p_{\parallel}=\left( \frac{\mu_{\parallel}}{kT}\right) ^2; \quad p_{\bot}=\frac{1}{2}\left( \frac{\mu_{\bot}}{kT}\right) ^2; \quad q = \frac{\Delta \alpha}{kT}.
  \label{eq:ck0}
\end{equation}
\(S^{e}(E)\) is the equilibrium order parameter reached after applying the field E for a very long time, \(C_K^0\) is the DC-field Kerr constant of the colloidal dispersion, and the parameters \(p_{\parallel},p_{\bot}\) and q describe the contributions of the permanent and induced dipole moments to the electric energy of the particle.

When the field is switched off, due to the rotational diffusion of the particles, the birefringence decays with time, following a simple exponential law:
\begin{equation}
  \Delta_n^{\textrm{off}}(t)=\Delta n^e(E) \exp \left(-6D_{\bot}^rt \right),
\end{equation}
where the relaxation time \(\tau_{\mathrm{off}}=1/(6D_{\bot}^r)\) corresponds to the rotational diffusion related to the second-rank tensor \(\bm{\alpha^{\textrm{opt}}}\).
The rise behavior of the induced birefringence when the field is switched on involves two more relaxation times: \( 1/(2D^r_{\bot}) \) and \(1/(D^r_{\bot}+D^r_{\parallel})\), which are respectively related to the longitudinal and transverse components of the permanent dipole moment of the particle \cite{Tinoco1959} (see SI for more details). A simpler treatment of the data is possible when $D^r_{\parallel} \simeq D^r_{\bot}$ or when \(D^r_{\parallel} \gg D^r_{\bot} \). In both cases, only two exponentials are needed to fit the rise curve: 
\begin{equation}
	\Delta n^{on}(t)=\Delta n^e(E) \left[ 1-\frac{3\beta}{2(\beta+1)} \exp(-2D^r_{\bot}t) + \frac{\beta-2}{2(\beta+1)} 
	\exp(-6D^r_{\bot}t) \right],
	\label{eq:2decays}
\end{equation}
where \(\beta\) is respectively \((p_{\parallel}-p_{\bot})/q\) and \(p_{\parallel}/(q-p_{\bot})\) in each case.\\

With bursts of sinusoidal field with frequency \(f\) and amplitude \(E_0\), \(E(t)=E_0\cos (2\pi ft)\), the signal relaxes to a steady-state regime after a transient initial response. Then, the birefringence has two contributions: a stationary (DC) one, \(\Delta n ^{\textrm{st}}(f)\), and an AC one, \(\Delta n ^{\textrm{osc}}(f)\), oscillating at a frequency double of that of the field: 
\begin{equation}
	\Delta n(t)=\Delta n ^{\textrm{st}}(f)+\Delta n ^{\textrm{osc}}(f)\cos (4\pi ft-\delta (f)).
	\label{eq:nteb}
\end{equation}

The information about the permanent and induced dipoles and the rotational diffusion of the particles is contained in the frequency dependences of the amplitudes of the two components. (The phase shift \(\delta \) in eq. \ref{eq:nteb} cannot be exploited with our external-electrode setup.) These dependences have been analyzed in detail by Thurston and Bowling (Th-B) \cite{Thurston1969} for a particle with revolution symmetry and with permanent dipole oriented only along the symmetry axis, i.e. with \(p_{\bot}=0\) . The stationary response is given by:
\begin{equation}
	\Delta n ^{\textrm{st}}(f) = \Delta n ^{e}(E_{\textrm{rms}})\frac{1}{P+1}\left(P+\frac{1}{1+\left(\pi f /D_{\bot}^r \right)^2} \right)=C_K^{\textrm{st}}(f) E_\textrm{rms}^2.
	\label{eq:stat_resp}
\end{equation}
where \(E_\textrm{rms}=E_0/\sqrt{2}\) is the root mean square (rms) value of the field and the parameter, \(P=q/p_{\parallel}\) describes the relative weights of the induced and permanent dipoles.

The oscillating response is given by:
\begin{equation}
	\Delta n ^{\textrm{osc}}(f)=\Delta n^{e}(E_{\textrm{rms}})\left[1+\left(\frac{P}{P+1}\right)^2\left(\frac{\pi f}{D^r_{\bot}}\right)^2\right]^{\frac{1}{2}} \left[1+\left(\frac{\pi f}{D^r_{\bot}}\right)^2\right]^{-\frac{1}{2}}\left[1+\left(\frac{2\pi f}{3 D^r_{\bot}}\right)^2\right]^{-\frac{1}{2}}
	\label{eq:oscill_resp}
\end{equation}

 At very low frequency, \(\pi f /D_{\bot}^r \ll 1\), the particles always follow the field and \(\Delta n ^{\textrm{st}}(f)\) remains constant, on a low-frequency plateau, 
\begin{equation}
	\Delta n ^{\textrm{st}}(0) = \Delta n^{e}(E_{\textrm{rms}})=  \frac{1}{15}\Phi \Delta n^p (p_{\parallel} - p_{\bot} + q) E_{\textrm{rms}}^2=C_K^0 E_{\textrm{rms}}^2.
	\label{eq:lowf-plateau}
\end{equation}

We note that the equilibrium value of the birefringence has this simple form even when \(p_{\bot} \neq 0\). \cite{Thurston1969,Kalmykov2009a} 

Upon increasing frequency, the rotational relaxation of the particles takes place and they only partially follow the field. Through this process, the rms-value of the induced-dipole torque remains constant because it is quadratic in the field. In contrast, the permanent-dipole torque, which is linear in the field, decreases with increasing frequency and vanishes at \(\pi f /D_{\bot}^r \gg 1\). In that limit, \(\Delta n ^{\textrm{st}}(f) \)  reaches a new, high-frequency, plateau:
\begin{equation}
	\Delta n ^{\textrm{st}}(\infty) = \Delta n^{e}(E_{\textrm{rms}})\frac{P}{P+1}=  \frac{1}{15}\Phi \Delta n^p qE_{\textrm{rms}}^2=C_K^{\infty} E_{\textrm{rms}}^2,
	\label{eq:highf-plateau}
\end{equation}
which depends only on the polarizability of the particle. Therefore, the ratio of the two plateaus is given by:
\begin{equation}
	\frac{C_K^{0}}{C_K^{\infty}}=1+\frac{p_{\parallel}-p_{\bot}}{q},
	\label{eq:ratiock}
\end{equation}
and provides direct information about the relative importance of the permanent and induced dipoles and their orientation and anisotropy.

\section{Results and discussion}
\begin{figure}[htbp]
  \centering
  \includegraphics[width=0.95\textwidth]{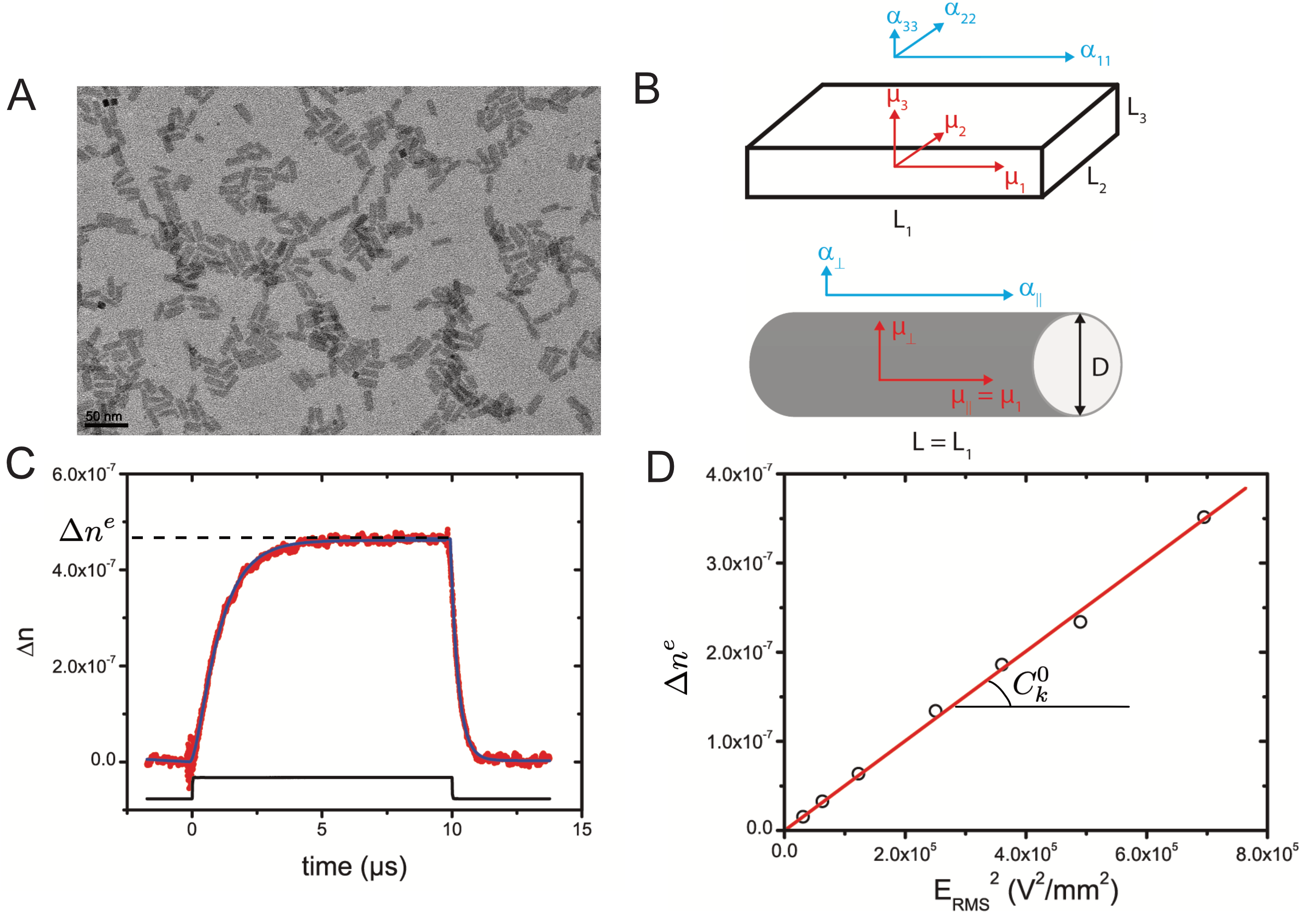}
  \caption{A) Transmission electron microscopy image of CdSe nanoplatelets. B) Schematic representation of a CdSe platelet, shown as a parallelepiped, approximated as a cylinder with revolution symmetry along the \textbf{1}-axis. The components of the polarizability tensor are shown in blue and those of the electric dipole moment are shown in red. C) Transient electric birefringence of a colloidal dispersion of isolated CdSe platelets in hexane ($\Phi$ = 5.9\(\times\)10$^{-4}$). The black line shows the evolution with time of the applied field and the red circles are the data points of \(\Delta n(t)\) induced by the short DC pulse (E = 950 V/mm). The blue lines are fits with the T-Y model (see equation \ref{eq:2decays}). D) Equilibrium value of the birefringence \(\Delta n^e\) as a function of the field squared, showing a linear behavior, with slope \(C_K^0\), the Kerr constant.}
  \label{fig:1}
\end{figure}
We synthesized CdSe nanoplatelets as described in the experimental section. Transmission electron microscopy images (Fig. 1.A) show that the NPL are parallelepipeds with “bare” (i.e. without the ligand brush) mean dimensions L$_1^b$=20 nm, L$_2^b$ = 9 nm, and L$_3^b$ = 1.5 nm. They are initially coated with the entropic ligand 2-hexyldecanoate which provides them with a longer colloidal stability \cite{Yang2016}. We assume that the short branched ligand measures 1.2 nm and is evenly located all over the particle. The addition of oleic acid slowly destabilizes the colloidal suspension as the CdSe platelets stack in wires whose average length increases with time over several months \cite{Jana2015,Jana2016}. We therefore studied the TEB signal not only of the initial colloidal suspension of “isolated” platelets (i.e. of independent particles) but also of the slowly growing particle stacks. We first describe the results on isolated NPL and show what information can be extracted from these experiments. A second part of the paper deals with stacks of NPL that form over a few months upon addition of oleic acid. 

\subsection{Isolated Platelets}
In the first few days after dispersion of the colloid, the TEB response is that expected for small isolated particles. The typical response to a short square pulse of DC field is shown on Fig. 1.C.  When the field is switched on, there is a fast increase of the induced birefringence, \(\Delta n(t)\), which then levels at a small positive equilibrium value, \(\Delta n^{e}(E)\). This demonstrates that the NPL align in the field. When the field is switched off, the birefringence decreases back to zero with an even faster relaxation time. As we will show, in fact, this behavior agrees well with the model of Tinoco and Yamaoka (T-Y) \cite{Tinoco1959} for the TEB of particles with cylindrical symmetry of the polarizability tensor and of the rotational diffusion coefficient.\\
In the rotational diffusion process, the particle and the ligand brush reorient together as a single rigid body that we call the “dressed” particle (i.e. including the ligand brush). Considering the brush thickness, the dressed particle dimensions along the three axes, \(L_i\), are respectively 22.4, 11.4 and 3.9 nm. The calculation of the rotational coefficients \(D^r_i\) around the axes i = \textbf{1}, \textbf{2}, \textbf{3}, of the dressed particle in hexane, using the Perrin formulae \cite{Perrin1934a,Perrin1936}, gives respectively $D_1 = 1.8 \times 10^6$ s$^{-1}$, $D_2 = 7 \times 10^5$ s$^{-1}$, $D_3 = 7.6 \times 10^5$ s$^{-1}$ (see SI for more details). Since the last two values are the closest, the best uniaxial approximation for the reorientation of an isolated CdSe platelet is a rod with length $L_{\parallel}=L_1$, diameter \(L_{\bot}=\sqrt{4L_2L_3/\pi}\) and approximate rotational diffusion constants $D_{\parallel}^r=1.8 \times 10^6$ s$^{-1}$ and $D_{\bot}^r=7.3 \times 10^5$ s$^{-1}$. The components of the dipole moment and polarizability tensor of the equivalent rod are then obtained from those of the platelet (Fig. 1B): \(\mu_{\parallel} = \mu_1; \mu_{\bot} = \sqrt{\mu_2^2+\mu_3^2}\) and \(\alpha_{\parallel}=\alpha_{11}\); \(\alpha_{\bot}=(\alpha_{22}+\alpha_{33})/2\). The Kerr constant of isolated platelets, deduced from the slope of the equilibrium value as a function of the field (Fig. 1.D.) is positive and very small, \(C_K^0=6.0 \times 10^{-19}\) m$^2$/V$^{2}$. Because both \(\Delta n^p\) and \(\Delta \alpha\) are positive (see supplementary information for detailed calculation) for the isolated particle, approximated as an effective rod, we deduce from the sign of  \(C_K^0\) and Eq. \ref{eq:deltane} that  \(p_{\parallel} - p_{\bot} + q > 0\), i.e. that the long axis of the NPL (the 1-axis of the equivalent rod) orients parallel to the field.\\
The best fit of the signal decay (Fig. 1C) gives $\tau_{\textrm{off}}=$ 0.30 $\mu$s and $D_{\bot}^r = 5.5 \times 10^5$ s$^{-1}$ which is in fair agreement with the previously estimated value of \(7.3 \times 10^5\) s\(^{-1}\). We note that this extremely short relaxation time is close to the time-resolution of our experimental setup and is therefore overestimated. Indeed, the deconvolution of the data from the instrumental function, separately measured, gives $D_{\bot}^r = 6.6 \times 10^5$ s$^{-1}$, which agrees better with the theoretical prediction.\\
For our CdSe particles, $D_{\parallel}^r \simeq 3D_{\bot}^r$, so that neither approximations required to describe the birefringence decay (Eq. \ref{eq:2decays}) holds true. However, in our case, a simple approach consists in comparing the two areas, \(I^{\textrm{on}}\) and \(I^{\textrm{off}}\), limited by the on- and off- curves (see SI for more details). The experimental value, \(\frac{I^{\textrm{on}}}{I^{\textrm{off}}} \simeq 3.57\), is much larger than 1, which shows the important contribution of the permanent dipoles to the birefringence. Indeed, in the opposite case where \(q \gg p_{\parallel}, p_{\bot} \), the induced birefringence is mainly due to the polarizability of the particle and the ratio is close to 1. With the numerical estimations of the diffusion coefficients, we obtain $p_{\parallel}-p_{\bot}/2 \simeq 5(q-p_{\bot}/2)$. Taking into account that \(p_{\parallel} - p_{\bot} + q>0\) (from the sign of the Kerr coefficient) and that the dipole moments are positive by definition, we obtain the inequalities \(p_{\parallel} > q > p_{\bot}/2 \geq 0\). This means that the most important contribution to the TEB of isolated particles comes from a large permanent dipole along the 1-axis of the platelet. We also note that q > 0, in good agreement with our rod-like approximation for the isolated CdSe platelet.\\

\begin{figure}[htbp]
  \centering
  \includegraphics[width=\textwidth]{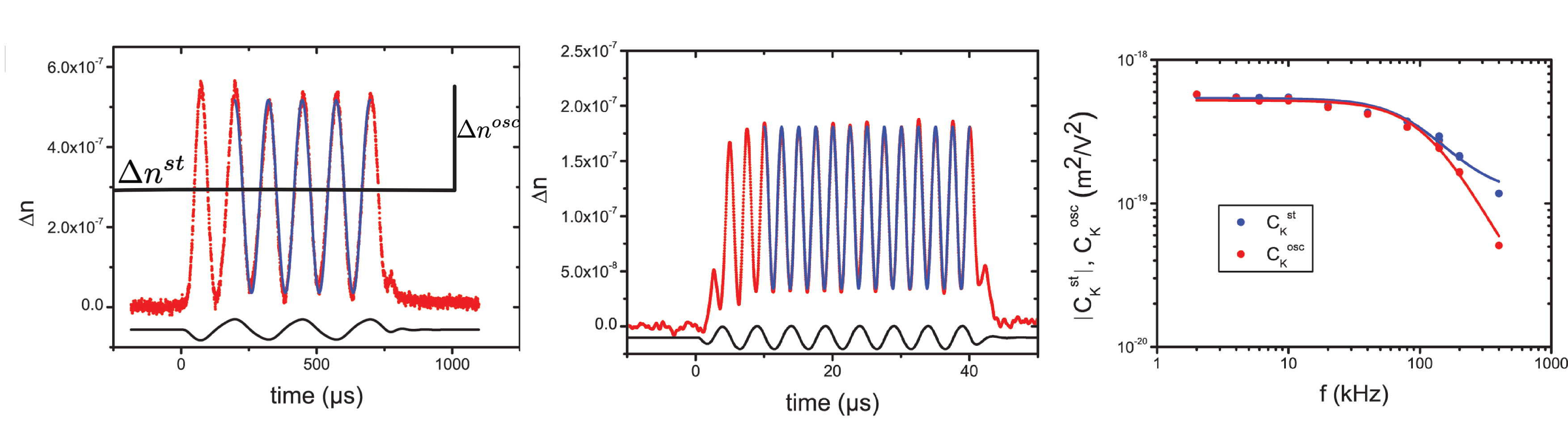}
  \caption{ A, B: \(\Delta n(t)\) induced by bursts of low- and high-frequency AC field (4 and 200 kHz). The blue lines are fits of the steady-state part of the curves with the Th-B model (equation \ref{eq:nteb}). C: Double-logarithmic plot, versus frequency, of the Kerr constants corresponding to the steady (blue circles) and oscillating (red circles) TEB contributions. The lines are fits of the data with the Th-B model (Eq. \ref{eq:stat_resp}, \ref{eq:oscill_resp})}.
  \label{fig:2}
\end{figure}

Typical TEB responses to sinusoidal bursts, measured at low and high field frequency, are displayed in Fig. 2.A) and 2.B). The induced birefringence is positive at all frequencies, showing that the major components of the permanent and induced dipoles are parallel. \cite{Thurston1969}

Fig. 2.C) shows the logarithmic plot, versus frequency, of the Kerr constants corresponding to the steady and oscillating TEB contributions. The continuous lines on the figure show the best fits of the experimental data with the Th-B model (Eq. \ref{eq:stat_resp}, \ref{eq:oscill_resp}). Qualitatively, both the steady and oscillating Kerr constants follow the expected trend, decreasing strongly at high frequencies. However, the fit is not quite satisfactory in the 20 – 80 kHz frequency range, suggesting the presence in this range of some additional relaxation process unrelated to rotational diffusion. \(\Delta n ^{\textrm{st}}(f)\) remains positive in the whole frequency range accessible with our set-up. It decreases by almost one order of magnitude but does not yet reach the second plateau, showing that \(C_K^{0}/C_K^{\infty} \geq 10 \). Since \(q=\Delta \alpha / (kT) >0\) for the isolated particle, approximated as a rod, we conclude, using Eq. \ref{eq:ratiock}, that \(p_{\parallel}>p_{\bot}\) and \((p_{\parallel}-p_{\bot})/q \geq 10\). This important results confirms the existence of a large permanent dipole moment parallel to the length of the particle, as already inferred from the slow birefringence rise induced by short square pulses. 
Therefore, to a very good approximation, the contribution of the induced dipoles to the TEB signal is negligible in front of that of the permanent dipoles, which greatly simplifies the following derivation of the absolute value of the dipoles of the CdSe platelets. 
For this purpose, one can use the low-field data acquired in the Kerr-regime, where the field-induced order is small and the birefringence is proportional to the square of the field. By neglecting the \(q\) term, we obtain from Eq. \ref{eq:ck0}:
\begin{equation}
	p_{\parallel}-p_{\bot}=(\mu_{\parallel}^2-\mu_{\bot}^2/2)/(kT)^2\simeq \frac{15}{\Phi \Delta n^p}C^0_K.
\end{equation}
From the measurement of \(C_K^0=6.0\times 10^{-19}\) m$^2$/V$^2$, the volume fraction $\Phi=5.9 \times 10^{-4}$ known through the absorption measurement \cite{Yeltik2015} and the calculated specific birefringence \(\Delta n^p=0.39 \) (see SI), we obtain the effective value \(\sqrt{\mu^2_{\parallel}-(1/2)\mu^2_{\bot}} \simeq 245 D\) for the dipole moment of the colloidal particles. 

From these experiments, we demonstrate that the dipolar term is larger than the polarizability term and that the response to the field is mainly due to the dipolar component. Moreover, the dipolar component along the largest dimension of the NPL is much larger than the one perpendicular. Finally, we can extract from the Th-B model a lower bound of this component of the dipole: $\mu_{\parallel}$ is larger than 245 D.  

\subsection{Nanoplatelet stacks}
As mentioned above, we also studied the electro-optic behavior of assemblies of NPL. It is well known that upon the addition of oleic acid, NPL slowly assemble into stacks \cite{Jana2015} whose geometry and dynamics are expected to be very different from those of the NPL alone. For example, due to their large dimensions, they relax more slowly when the electric field is switched off. 
\begin{figure}[htbp]
  \centering
  \includegraphics[width=0.5\textwidth]{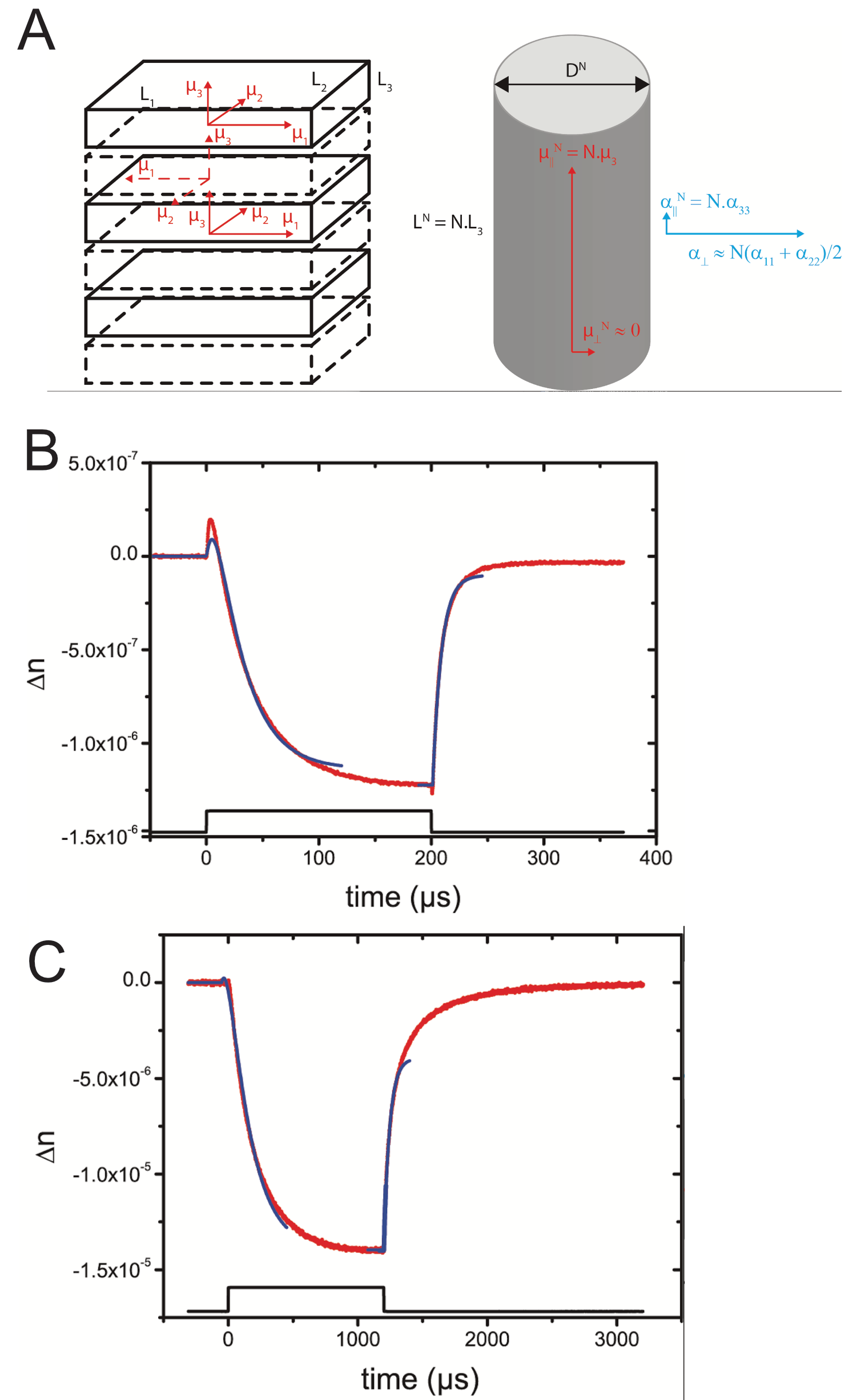}
  \caption{A) A stack of platelets is approximated as a cylinder (in grey) with revolution symmetry along the 3-axis. The subscripts \(\parallel\) and \(\bot\) refer to the orientation with respect to the revolution symmetry axis. The components of the polarizability tensor are shown in blue and those of the electric dipole moment are shown in red. On the left, the stack is made of platelets whose in-plane components of the dipole moment point alternatively in one direction (solid line) or the opposite one (dashed line). B) TEB of short stacks (25 days of aging) submitted to DC pulses (E = 950 V/mm).  C) TEB of long stacks (105 days of aging) submitted to DC pulses, (E = 900 V/mm). In B) and C), the blue lines show fits of the data with the T-Y model (see equation 7) and, due to polydispersity effects, only the initial regions of the curves are fitted.}
  \label{fig:3}
\end{figure}

At t=0, we added oleic acid to the dispersion and followed the optical response as a function of time. We expect the long-chain acid addition to trigger the slow destabilization of the NPL and the formation of stacks. In the first few days after dispersion of the colloid, the TEB response remains that expected for small isolated particles.
A few days after the injection of oleic acid into the dispersion of NPL, the TEB signal started to evolve gradually from the fast response, with small amplitude, of isolated particles described previously to a slower response with larger amplitude, showing the occurrence of particle stacking. Moreover, the induced birefringence measured with short pulses or under bursts of low-frequency field changed its sign and became \textit{negative}, which indicates a drastic change in the geometry of the reorienting objects.\\

Fig. 3.B) shows the TEB signal of the short stacks (St1) that appear at an early stage of particle stacking (after 25 days). To interpret this TEB signal, the geometry and the physical properties of the stacks must first be discussed. Previous x-ray scattering and electron microscopy studies \cite{Jana2015} have shown that stacking takes place along the normal to the platelets, i.e. along their 3-axis (Fig. 3.A). Moreover, the condition of minimum electrostatic energy of the stack imposes that adjacent particles have parallel \(\mu_3\) components of their dipole moment, but anti-parallel \(\mu_1\) and \(\mu_2\) components. Supposing that the particles are densely stacked, a stack of N particles will have the dimensions $L_3^N=N \times L_3$, $L_1^N=L_1$ and $L_2^N=L_2$. Because $L_3^N$ rapidly increases with $N$, for $N>10$, the equivalent shape of the stack transforms to a rod elongated along the 3-axis. Even the smallest stacks (St1, after 25 days) that we investigated electro-optically have \(N \simeq 23\) (as shown in the following) and are moderately long rods, with rotational diffusion constants \((D^r_i)^N\) respectively of \(1.6 \times 10^4\) s\(^{-1}\), \(1.8 \times 10^4 \) s\(^{-1}\), \(1.1 \times 10^5\) s\(^{-1}\) (see supplementary information for details). Therefore, for the interpretation of the electro-optic data, we approximate the stack of N particles as a rod of length $L_{\parallel}^N=N\times L_3$ and diameter $L_{\bot}^N=\sqrt{4L_1L_2/\pi}$ with $(D^r_{\parallel})^N \gg (D^r_{\bot})^N$. For the dipole moment of the equivalent rod of the N-stack, we obtain a large longitudinal component, $\mu_{\parallel}^N = N \times \mu_3$ because the $\mu_3$ components add in the stack. On the contrary, the transverse component is very small: it is either $\mu_{\bot}^N=\sqrt{\mu_1^2+\mu_2^2}$ for odd $N$ or it vanishes for even $N$, due to the alternating 180$^{\circ}$ rotation around the \textbf{3}-axis of the particles in the stack.
In contrast, the second-rank polarizability tensor, $\bm{\alpha}$, is invariant upon 180$^{\circ}$ rotation of the particle. Therefore, assuming that the CdSe particle cores in the stack are electrically insulated by their organic ligand brush (whose physical properties are similar to those of hexane), all three components of $\bm{\alpha}$ should be additive (see SI for more details).\\
When the dressed particles are stacked, the polarizability density per unit volume and then the specific refractive indices remain unchanged after stacking: \((n^p_i)^N=n^p_i\) because the polarizability of a stack depends linearly on N and the total number of particles in the sample remains constant during stacking. However, contrary to the case of an isolated particle, the stack behaves hydrodynamically as a rod with revolution symmetry around the 3-axis. This leads to \((n^p_{\parallel})^N=n^p_3=1.85\), \((n^p_{\bot})^N=\sqrt{(n_1^p)^2+(n_2^p)^2/2}=2.44\), and \((\Delta n^p)^N=-0.58\), a \textit{negative} value. Consequently, the equivalent optical polarizability tensor of the rod-like stack is oblate, i.e. as expected for a disk. This very unusual feature is due to the complex stack structure: the stack rotates as a rigid rod-like body but its optical response is that of a disk-like polarizable particle. In a similar way, based on the different known polarizability mechanisms, we expect that the electric polarizability at low frequency of the stacks is also additive, leading to: \(\alpha_{\parallel}^N=N \times \alpha_{33}\) and \( \alpha_{\bot}^N \simeq N\times(\alpha_{11}+\alpha_{22})/2\) (see SI for details).\\
Taking into account the effective rod geometry of the stacks, the negative sign of the induced birefringence (Fig. 3.B) indicates that, unlike the case of isolated particles, the major components of the permanent and induced dipoles of the stack are perpendicular. The TEB decay is not exponential, suggesting that the system is polydisperse. Moreover, the small and very fast overshoot at the beginning of the decay curve is due to isolated particles that still remain in the dispersion and that give a positive contribution to the birefringence. The best decay fit provides \(\tau = 9.3 \, \mu\)s and \(D^r_{\bot}=1.8 \times 10^4\) s\(^{-1}\), i.e. the rotational diffusion of the short stacks is about 30 times slower than for isolated particles. The T-Y fit of the TEB rise is reasonably good but it deviates from the data at both ends of the curve due to the polydispersity and the presence of isolated particles. The best fit parameters are \(\beta=(p_{\parallel}-p_{\bot})/q \simeq -4\) and $C_K^0=-1.3 \times 10^{-18}$ m$^2$/V$^2$. Since the polarizability anisotropy \(\Delta \alpha\) (and \(q=\Delta \alpha/(kT)\)) is negative for the rod-like stacks, the negative sign of \(\beta\) and \(C_K^0\) indicates that the permanent dipole of the stack is parallel to its long axis (i.e. the stacking axis). \\
The TEB response of longer stacks under DC pulses (St4, after 105 days) is presented on Fig. 3.C). The experimental curves deviate strongly from the theoretical predictions and the fit with the T-Y model is good only in the initial regions. This behavior is most probably due to the stack polydispersity. The induced birefringence is again negative and is much stronger than for the shorter stacks (St1). The lack of overshoot of the decay curve suggests the absence of isolated particles. The best decay fit gives \(\tau=57 \mu\)s and \(D^r_{\bot}=2.9 \times 10^3\) s\(^{-1}\), i.e. the rotational diffusion of the long stacks is about 200 times slower than for isolated particles. The T-Y fit of the TEB rise is reasonable only in the first 400 $\mu$s of the signal, which is probably again due to the large stack polydispersity. The best fit parameters are \(\beta=p_{\parallel}/q \simeq -9 \) and \(C_K^0=-1.6 \times 10^{-17}\)m$^2$/V$^2$, showing again a large and dominant contribution of the permanent dipole along the stack long axis. We note also that when \(p_{\parallel} \gg \lvert q \lvert \), which is actually the case for the stacks, the T-Y fit is rather indiscriminative for the precise value of \(\beta\). Therefore, the previous results, \(\beta \simeq -9\), is only qualitative and just means that the TEB response is dominated by the large permanent dipole of the stack.\\
Qualitatively, these conclusions are confirmed by the TEB signal (Fig. 4.A-C) of the same long stacks under bursts of low-, medium-, and high-frequency AC field. The steady component of the induced birefringence is negative at low frequency, vanishes at around 14 kHz and is positive at higher frequency. This behavior shows clearly that both the electrical, \(\bm{\alpha}\), and the optical, \(\bm{\alpha}^{\mathrm{opt}}\), polarisabilities of the stack are oblate tensors (\(\Delta \alpha = \alpha_{\parallel}-\alpha_{\bot} < 0\)) and that the permanent dipole of the stack is parallel to its long axis.\\

The same salient features of the TEB behavior were observed throughout the growth of the CdSe platelet stacks: (i) the rise and decay times increased because of the decrease of the rotational diffusion coefficient \( D^r_{\bot}\); (ii) the Kerr constant at low frequency, \(C_K^0\), remained negative and its absolute value increased with aging time (and hence with the stack length); (iii) the Kerr constant at high frequency, \(C_K^{\infty}\), remained positive and much smaller than \(\mid C_K^0 \mid\) and (iv)  the frequency \(f_0\) defined by \(C_K^{\mathrm{st}}(f_0)=0, \) decreased with increasing stack length because of the decrease in \( D^r_{\bot}\).

\begin{figure}[htbp]
  \centering
  \includegraphics[width=\textwidth]{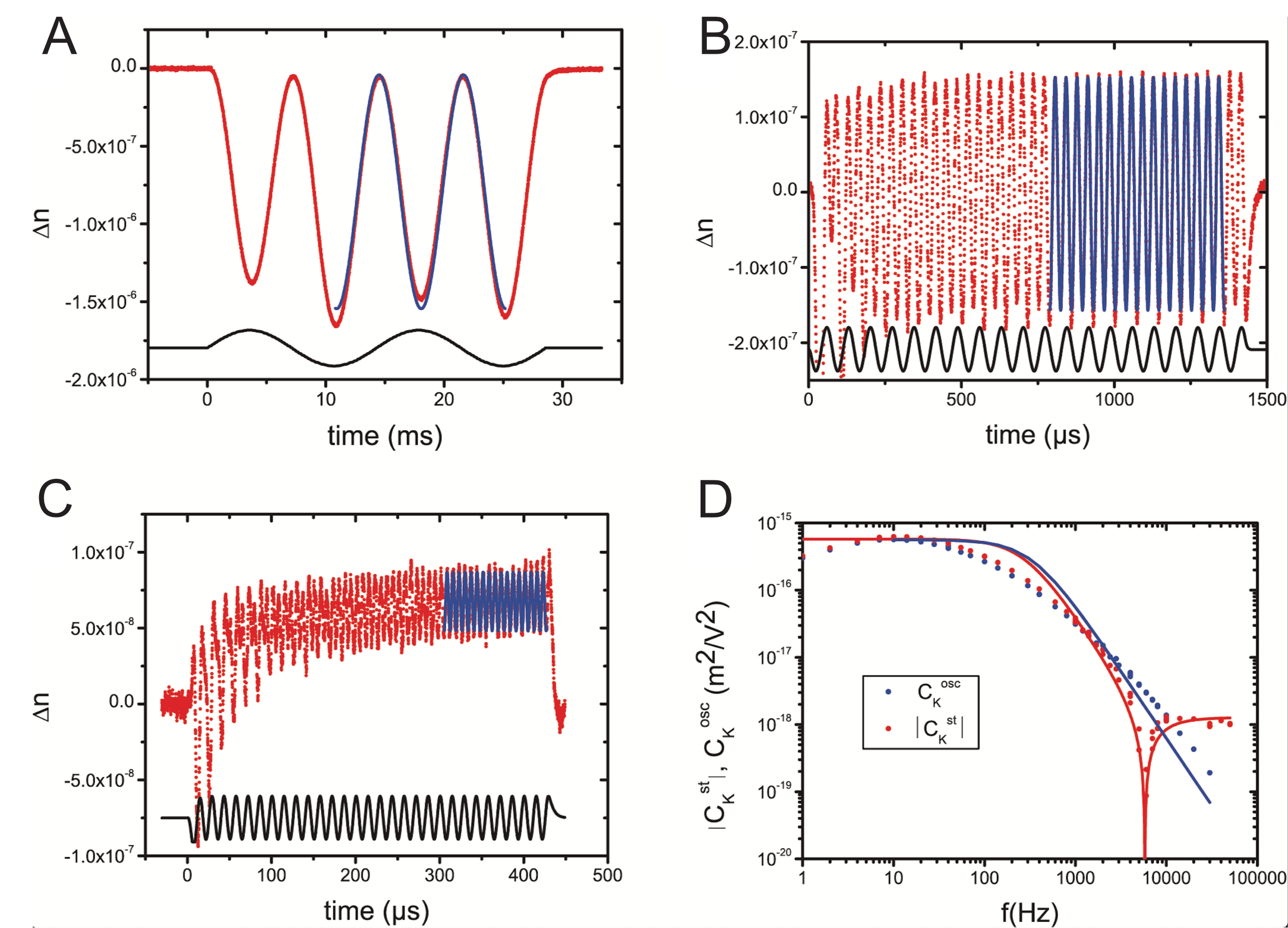}
  \caption{A, B, and C: TEB of the long stacks (105 days) submitted to bursts of low-, medium-, and high-frequency AC field: 70 Hz, 14 kHz and 70 kHz respectively. The blue lines are the fits of the relaxed part of the curves with the Th-B model (Eq. \ref{eq:nteb}). D: Double-logarithmic plot, versus frequency, of the Kerr constants corresponding to the steady (red circles) and oscillating (blue circles) TEB contributions for the largest stacks (415 days of aging). The lines are fits of the data with the Th-B model (Eq. \ref{eq:stat_resp}, \ref{eq:oscill_resp}).}
  \label{fig:4}
\end{figure}

Fig. 4.D. displays the frequency dependence of the steady (\(C_K^{\textrm{st}}\)) and oscillating (\(C_K^{\textrm{osc}}\)) Kerr constants of the largest stacks (St5, after 415 days). Qualitatively, the behavior of the two curves follows the trend expected for a rod with \(\Delta \alpha < 0\), \(\Delta \alpha^{\mathrm{opt}} < 0\) and a large longitudinal permanent dipole moment. However, the theoretical model describes the experimental curve better or worse in the different frequency domains. Below \(f=10\) Hz, both \(C_K^{\mathrm{st}}\) and \(C_K^{\textrm{osc}}\) decrease instead of remaining constant. This artifact is simply due to our external-electrodes technique for applying the field. Indeed, the field penetrating in the sample at these low frequencies is partially screened by the conductive charges in the solvent. However, despite this difficulty, the low-frequency plateau is well-enough pronounced, corresponding to a Kerr coefficient \(C_K^0 = -5.8 \times 10^{-16}\) m$^2$/V$^2$, 400 times larger than for the shortest stacks (St1). The Th-B model does not describe the experimental curves well in the region between 30 and 800 Hz. This other discrepancy may be due to the large polydispersity of the stacks or to the fact that the Th-B model assumes that there is no other relaxation process than rotational diffusion in the frequency range under study. However, at higher frequencies, both around the sign-inversion frequency of \(C_K^{\textrm{st}}\), \(f_0=5.9\) kHz, and above, on the high-frequency plateau, the theory is in good agreement with the experimental curve. The Th-B fit of \(C_K^{\textrm{st}}\) in this region provides \(D^r_{\bot}=1020\) s\(^{-1}\), \(\beta=\) -330 and \(C_K^0=-5.8 \times 10^{-16}\) m$^2$/V$^2$, indicating that the strong induced birefringence is due to the huge permanent dipole moment along the long axis of the stack. The main experimental results for stacks of different ages are presented in Table I. The values of \(C_K^0\) obtained from the sinus bursts are very close to the pulse values, but are slightly more dispersed for the small stacks due to their weak TEB signal. The results for \(\beta=p_{\parallel}/q\) and \(D^r_{\bot}\) are those obtained from the value of the frequency \(f_0\) at which \(C_K^{\textrm{st}}\) changes sign and from the Th-B fit of \(C_K^{\textrm{st}}(f)\) in the vicinity of \(f_0\). Actually, these values are less influenced by the stack polydispersity and non-rotational relaxation processes.

\begin{table}[htbp]
\begin{center}
\begin{tabular}{|c|c|c|c|c|c|c|c|c|}
\hline
	& 	&   \multicolumn{3}{|c|}{Experimental data} & \multicolumn{4}{|c|}{Results} \\
	\hline
	Experiment & Age & \(C_K^0\) & \(D^r_{\bot}\) & \((p_{\parallel}-p_{\bot})/q\) & \((p_{\parallel}-p_{\bot})\) & dipole & S & N \\
	           & (Days)  & (10$^{-18}$m$^2$/V$^2$) & (10$^3$ Hz) &   &   (10$^{-14}$m$^2$/V$^2$)& (D)  &    &    \\
	           &         &   a)    &   b)    &c)&c)&d)&e)&f)\\
	           \hline
	Stacks 1  &	25	&- 1.3	& 16.1	&- 11.5	&8.1	&350	&0.0059	&23\\
	\hline
	Stacks 2  &	35	&- 6.0	& 13.9	&- 17.7	&38	    &750	&0.027	&27\\
	\hline
	Stacks 3  &	62	&- 11.2	& 10.6	&- 38.4	&70	    &1030	&0.051	&32\\
	\hline
	Stacks 4  &	105 &- 15.9	& 4.13	&- 119	&99	    &1230	&0.073	&46\\
	\hline
	Stacks 5  &	415 &- 500	& 1.02	&- 330	&3120	&6900	&0.49	&92\\
	\hline
	Stacks 5  & 415 & -500  &       & -16.1 &3470   &7260   & 0.49   &  \\
	Saturation & & g) & & g) & g) &g) & g) & \\
	\hline 
\end{tabular}
\end{center}
\caption{Electro-optic properties of CdSe platelet stacks directly measured (experimental data) or deduced from data interpretation (results). \textbf{a)} Kerr constant calculated from the  DC pulses data; \textbf{b)} Rotational diffusion coefficient; the values are calculated from the sign-inversion frequency \(f_0\); \textbf{c)}	Ratio of the contributions to the TEB from the dipole moment and polarizability: \(p_{\parallel}-p_{\bot} = \left(\mu_{\parallel}^2-\mu_{\bot}^2/2\right)/(kT)^2, q=\Delta \alpha/(kT)\); the values are calculated from the sign-inversion frequency \(f_0\); \textbf{d)} dipole: \(\sqrt{\mu_{\parallel}^2-\mu_{\bot}^2/2}\); \textbf{e)} Orientational order parameter measured at \(E=1 V/\mu m, S(E)=\Delta n(E)/\Delta n^{\mathrm{sat}}\); \textbf{f)} Number of particles in the stack calculated by comparing the experimental and theoretical D$^r$ values; \textbf{g)} Values obtained from the fit of the saturation curve.}
\end{table}

\subsection{Saturation of the birefringence for long stacks}
\begin{figure}[htbp]
  \centering
  \includegraphics[width=0.5\textwidth]{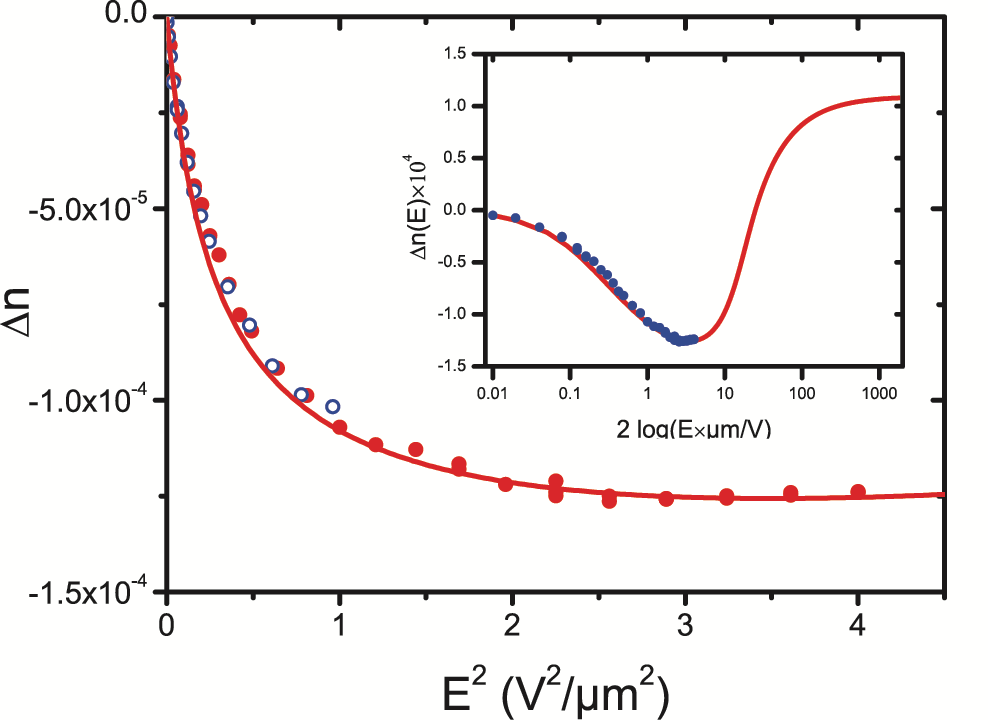}
  \caption{Saturation of the induced birefringence at high field (10 ms long DC pulses) for the largest stacks (415 days of aging). The blue and red circles correspond to two independent experiments (see text). The red line is the best fit with the theoretically predicted behavior (see text). The inset shows the same information in log-lin representation. }
  \label{fig:saturation}
\end{figure}
At large field, away from the Kerr regime, the induced order should be strong enough to lead to the saturation of the TEB signal. This is indeed observed with the largest stacks, for which \( \Delta n (E)\) significantly deviates from the $E^2$ law, even though it does not reach complete saturation for the fields (\( E \leq 1 \) V/\(\mu\)m) accessible with our set-up in usual conditions (Fig. 5). However, using our “double-field” trick (see experimental section), we managed to apply inside the sample, in a transient way, fields up to 2 V/$\mu$m and, therefore, reach the complete saturation of \( \Delta n (E)\) (figure \ref{fig:saturation}). 
The usual treatment of this kind of TEB data \cite{Shah1963,OKonski1959} based on the series expansion in \(E^2\) of \(S(E)\) up to the \(E^4\) term, works well for the case \((p_{\parallel}-p_{\bot})/q \geq 0\), when the \(\Delta n(E)\) curve is monotonous. However, for large negative \((p_{\parallel}-p_{\bot})/q \) ratios, as in our case, the series converges too slowly and a large number of terms should be included, making this approach impractical. Therefore, we fitted our experimental data with the function \(\Delta n(E) = \Delta n^{\textrm{sat}}S(E)\), where the order parameter \(S(E)\) is calculated numerically, assuming \(\mu_{\bot}=0\). The fit of the data is excellent and provides values presented in the last row of Table 1.


We note that these values of \( \mu_{\parallel} \), \(\Delta \alpha\) and  \(C_K^0\) are self-consistent and independent of the Kerr-regime measurements under DC pulses and AC bursts. 
The maximum value of the order parameter is \(S^{\mathrm{max}} = S(E=1.8)\) V/$\mu$m) = 0.57 which corresponds to a “saturated” birefringence value of \(\Delta n^{\textrm{sat}}=\Delta n(S=1)=-2.19 \times 10^{-4}\). This value is of the same order of magnitude but smaller than the one estimated from the measured volume fraction and the calculated specific birefringence of the stack (\(\Delta n^{\textrm{sat}} = \Phi \Delta n^p=-3.14 \times 10^{-4} \)). However, the former value is directly derived from a self-consistent experiment and is not based on any approximation. Therefore, we used it to calculate \( \mu_{\parallel}\) and \(\Delta \alpha\) from the experiments on the stacks at low field. \\

Altogether, our TEB experiments provide a measurement of the dipole component \(\mu_{3}^N\) for different nanoplatelet stacks with increasing size \(N\), ranging from 350 D for the first and smallest stacks to 7260 D for the largest ones after more than a year (Table 1). These huge values are the physical origin of the very important TEB signal that we measured after addition of oleic acid. We stress that these results are very complementary with those obtained with  isolated NPL which provided a value of \(\mu_{\parallel}\). With the two sets of measurements, we can estimate the two orthogonal dipolar components if we manage to extract the value of \(\mu_{\bot}\) for individual platelets from our measurements of \(\mu_{\bot}^N\) of stacks. We describe a method to do so in the following paragraph. 

\subsection{Growth of nanoplatelet stacks with aging time}
\begin{figure}[htbp]
	\centering
	\includegraphics[width=0.5\textwidth]{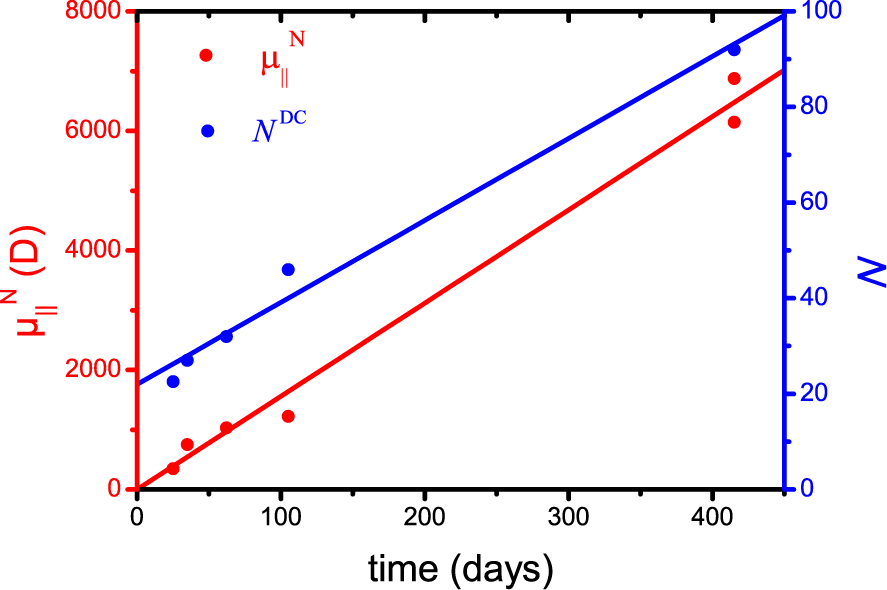}
	\caption{Evolution with time of the component of the electric dipole moment (red symbols) parallel to the main axis of the stacks and of the number of platelets per stack determined by modelling the TEB of stacks submitted to DC pulses (blue symbols).}
	\label{fig:6}
\end{figure}
The approximately linear dependence of \(\mu_{\parallel}^N\) on aging time (Figure 6) reflects the increase in average number of particles \(N_{\textrm{av}}(t)\) and in length of the stack with time. \(N_{\textrm{av}}(t)\)
can be estimated by comparing the values of \(D^r_{\bot}\) with the values calculated numerically \cite{Perrin1934a,Perrin1936} for different values of \(N_{\textrm{av}}\). We calculate the rotational diffusion constants \(D_i^r\) (i=1,2,3), using the known dimensions \(L_i\) of the dressed platelet and approximating the stack as a rigid biaxial ellipsoid with the same volume and axial ratios as those of a rectangular prism of dimensions \(L_1, L_2, N_{\textrm{av}} \times L_3 \). The calculated value of \(D^r_{\bot}=(D^r_1+D^r_2)/2\) is plotted versus \(N_{\textrm{av}}\) in figure S2. The \(D^r_{\bot} \) data was derived from the decay time of the TEB signals of the stacks, which were assumed to be monodisperse. The \(N_{\textrm{av}}(t)\) values, called \(N^{\textrm{DC}}\) were obtained by comparison of the experimental and calculated \(D^r_{\bot}\) and are plotted in Fig. 6 as a function of aging time. N is about 23 in the first experiment where the stacks were detected (25 days of aging) and is 3 – 4 times larger for the longest stacks. Despite the linear increase in time of both \(N^{\textrm{DC}}\) and \(\mu_{\parallel}^N\), these quantities are not really proportional (Fig. S2), as would be expected from the relation \(\mu_{\parallel}^N/N=\mu_3^1\). We explain in Supplementary Information how this discrepancy can arise from the presence of a fraction of isolated platelets coexisting with the stacks but slowly disappearing with aging time. This analysis leads us to the best estimate for the permanent dipole along the platelet normal: \(\mu_3 \simeq 80 D\).

\section{Discussion}
We thus find that CdSe NPL bear an important dipole whose magnitude is larger than 300 D. The in-plane component is larger than 245 D while the component along the thickness is around 80 D. These values are very large in comparison with previous direct measurements on dots and rods which yield typical dipoles ranging from 20 to 250 D. \cite{Greenwood2018} When scaled with the volume of the particles, this corresponds to 1.1 D/nm$^{-3}$, twice the value measured for wurtzite CdSe nanorods. \cite{Li2003} 
If we reason in the bulk, this is surprising in the first place since wurzite in known to be pyroelectric. The space group 6mm to which this structure belongs has a unique polar axis parallel to the 6-fold symmetry axis. In this direction, there is an alternation of short and long Cd-Se bonds. Furthermore, this axis is unique in the sense that it is not repeated by any symmetry element so that elementary dipolar moments add up. The zinc blende structure ($\bar{4}$3m) also displays 4 polar axes (in the <111> directions) but they are related by the $\bar{4}$ roto-inversion axis in such a way that the dipoles cancel. If the nanoparticles bear a permanent electric dipole moment, it is likely to be parallel to one of these directions. The basal planes of the CdSe NPL are of the {001} type and are neither parallel nor perpendicular to the <111> directions. Therefore, it is not a priori surprising that the permanent electric dipole has both parallel and perpendicular components with respect to the normal to the platelet. However, there is no symmetry reason for a particular <111> axis to be privileged compared with the other similar directions, so there must be another source of asymmetry in the system.\\

There are several phenomena that can induce a spontaneous symmetry breaking and make a permanent dipole emerge. First, a NPL has limited dimensions and cutting a crystal into a given shape can reduce its symmetry. If we consider that the NPL adopt a perfect parallelepipedic shape, this argument does not hold since there are still multiple polar axis along the diagonals of the parallelepiped. Though the zinc-blende CdSe structure presents an asymmetric alternation of long and short Cd-Se bonds along the <111> directions \cite{Khurgin1998}, any permanent electrical dipole along these particular axes will still be compensated within the perfect parallelepipedic shape. Hence, the symmetry reduction caused by cutting the crystal into a NPL can not explain alone the emergence of a permanent dipole. \\

However, the zinc-blende structure is piezoelectric. In the presence of stress, the bonds will deform and, depending on the orientation of the stress with respect to the crystalline structure, this will yield a dipole. For example, if the zinc-blende lattice is strained along the <111> axis a net polarization will appear in this direction since the Cd-Se bonds will be deformed in such a way that dipoles will not compensate anymore \cite{vonHippel1952}. Not all stresses will yield a polarization though. This is apparent from the shape of the piezoelectricity tensor which links stress and polarization. In the zinc-blende case, only three terms are non-zero \cite{Nye1984} and a deviatoric component to the stress is needed for polarization to emerge. For example, a simple deformation of the NPL along the direction perpendicular to the basal {001} plane is not enough to make a polar axis unique. The piezoelectric constant $e_{14}$ relates the polarization to the strain. \cite{Huong1998} For zinc-blende CdSe, it is estimated \cite{Berlincourt1963,Xin2007a} to be 0.2 C/m$^2$ . The dipole scaled to the volume that we measured (1.1 D/nm$^{-3}$) corresponds to a polarization of 3.63$\times$10$^{-3}$ C/m$^2$. Thus, a strain of only 1.8\% can explain our result with the literature value of $e_{14}$. It is well known that surface ligands induce stress at the surface of semi-conducting colloidal nanocrystals due to incompatibility between their preferred conformation and the lattice of the inorganic core. \cite{Meulenberg2004,Huxter2009} X-ray diffraction studies of CdSe spherical nanocrystals have shown that strain increases when the size of the nanocrystals decrease, reaching 0.5 \% for 2.2 nm CdSe nanoparticles. In the case of NPL, their even smaller thickness and their high ligand density \cite{Singh2018} are likely to generate larger strains. Ligand exchange from the native oleic acid to phosphonic acid or thiols has been shown to distort the crystal lattice significantly with relative variations of lattice parameters which could reach 4\%. \cite{Antanovich2017} These important strains are consistent with previous studies which have shown that CdSe NPL could adopt various curved conformations depending on the surface ligand and their crystallographic structure. \cite{Bouet2013a,Hutter2014,Jana2017} Due to the very thin nature of the NPL, even the small stress exerted by the ligands at their surface can result in large deformations. Atomic arrangements are modified by the surface stress and depart from their highly symmetric configurations. Consequently, the physical origin of the dipole could be the stress imposed by the organic ligands at the surface of the NPL. \\

We now discuss the consequences of the presence of a large permanent dipole in CdSe nanoplatelets. Such an important permanent dipole moment will affect the colloidal interactions between NPL and strongly impact their colloidal stability in suspension. \cite{Israelachvili2010} With the particle dimensions and the values of the components of the dipole moment that we derived above, an order of magnitude of this dipolar interaction energy can be estimated for different relative orientations of two platelets (keeping in mind that the finite size of the dipoles may not be neglected in front of their separation). The largest attraction energy, of about -3 kT, is found for two stacked platelets at contact (i.e. at 4 nm separation) when the $\mu_3$ components are in line and the $\mu_1$ components are anti-parallel. However, at room temperature, thermal averaging of the relative orientations of the dipoles should also be considered, a process leading to the Keesom interactions for freely-rotating point-like dipoles. Thermal fluctuations will also induce deviations from the ideal stacked configuration, resulting in an increase of the average separation between platelets. This thermal averaging will sharply decrease the magnitude of the interaction energy since the potential strongly depends on the platelet separation. This reasoning may qualitatively explain the marginal colloidal stability of CdSe nanoplatelets in hexane but a more rigorous statistical physics treatment of this question is required to reach a more quantitative description. \\

A large ground state dipole should also impact the optical properties of CdSe NPL. By breaking the inversion symmetry of the NPL, the internal electric field will mix odd and even quantum states. \cite{Schmidt1997} This should be visible in the difference between one photon and two-photon absorption spectra at low temperature and parity-forbidden transitions should be allowed. Such effects have already been shown to occur for CdSe spherical nanocrystals \cite{Schmidt1997} but their relevance for nanoplatelets is still to be assessed.

\section{Conclusion}
Using transient electrical birefringence on dispersion of CdSe NPL and their self-assembled stacks, we demonstrated that these nanoparticle bear an important permanent dipole larger than 300 D with components perpendicular (> 245 D) and parallel ($\simeq 80 D$) to the NPL normal. This corresponds to a very large polarization, almost twice larger than what has been previously observed in wurtzite nanoparticles though the zinc-blende structure is not polar. The dipole could arise from deformation of the crystalline lattice from its cubic structure due to ligand induce surface stress. Variation of the particle thickness might help rationalizing these results further. These results have important implications on the self-assembly of NPL into larger scale structures. It also highlights that an electric field could be used to orient very efficiently NPL in space to harness their outstanding anisotropic optical properties and their directed emission \cite{Scott2017}. Finally, this ground state dipole should be taken into account in order to understand 2-photon and Stark spectra at low temperatures. 

\section{Experimental}

\subsection{Synthesis and purification of CdSe nanoplatelets}
All chemical were purchased at Sigma-Aldrich. \\
\textbf{Synthesis of cadmium oleate}\\
40 mmol of sodium oleate is dissolved in 200 ml of ethanol and 50 ml H$_2$O mixture and stirred for 30 min at 60-70$^{\circ}$C until a clear transparent solution is obtained. The solution is then cooled to around ~40$^{\circ}$C . In another beaker 20 mmol of cadmium nitrate is dissolved in 50-60 ml of ethanol. This solution is slowly added to the Na-Oleate solution with constant stirring. After complete addition the mixture is kept stirring for another 30 minutes. A white precipitate is formed and the supernatant is discarded. Fresh ethanol is added and the precipitate is retrieved after a centrifugation at 3000 rpm for 5 min. The white product is washed 3/4 times by hot ethanol and finally washed with hot methanol. The final product is kept under vacuum overnight to dry. It should have the aspect of a white slightly sticky powder. \\

\textbf{Synthesis and purification of the NPLS}\\
404 mg of cadmium oleate, 27 mg of selenium powder (100 mesh), and 25 mL of octadecene (ODE, 90\%) were introduced into a 50 ml three-neck round bottom flask, equipped with a septum, a temperature controller and a condenser, and were kept under vacuum for 30 minutes. Afterwards, the flask was purged with argon and the temperature was set to 240$^{\circ}$C. At 180-190 $^{\circ}$C, the selenium started to dissolve and the solution turned clear yellow. When the temperature reached 205 ${\circ}$C, the septum was withdrawn and 140 mg of cadmium acetate (Cd(OAc)$_2$, 2H$_2$O,  Aldrich) was swiftly added into the flask. After the temperature reached 240$^{\circ}$C, the reaction continued for 12 minutes and 1 mL of oleic acid was injected at the end. The flask was immediately cooled down to room temperature. At this stage, the reaction product was a mixture of 5 monolayers (ML) NPL, a few 3ML NPL and quantum dots in solution. The 5 ML NPL were collected using size-selective precipitation by addition of ethanol and re-dispersion in 3mL of hexane. 50 uL of 2-hexyldecanoic acid were added. After 45 minutes, they were precipitated with 15 mL of acetone and centrifuged at 4000 rpm. The clear supernatant was discarded and the entire operation was repeated again.
To finish, the platelets were re-dispersed in hexane.

\subsection{Transient electric birefringence}
The experimental set-up for the TEB measurements was previously described in detail \cite{Dozov2011b,Paineau2012,Paineau2012a}. It is mostly inspired by classic TEB experiments \cite{OKonski1959}, except for one important modification: instead of the classic Kerr cell, with electrodes immersed in the liquid and long light-path (several centimeters) of the probe beam, the sample in our case was contained in a flame-sealed cylindrical glass capillary of diameter D = 1 mm. The electric field was applied parallel to the capillary axis by a pair of external electrodes (2 mm apart) placed directly on the outer surface of the capillary wall. The voltage applied to the electrodes was either as bursts of sinusoidal alternating current (AC) voltage (from 1 to 10$^4$ periods in one burst) with variable frequency f, ranging from 1 Hz to 400 kHz, or as short direct current (DC) pulses (duration $\tau_{imp}$ from 10 $\mu$s to 10 ms). The numerical simulation of the field penetration into the capillary \cite{Dozov2011b,Antonova2012} shows that the field inside the colloidal dispersion is uniform and that the screening losses due to accumulation of charges on the inner side of the capillary wall are negligible at high enough frequency (here, f > 10 Hz). \\

Low-voltage (< 10 V) AC bursts and DC pulses with the required repetition rate were generated by an Arbitrary Waveform Generator (TGA 1241, TTi) and sent to an amplifying block. This block consisted of several different instruments, depending on the required voltage amplitude, U,  and response time of the amplifier, $\tau_r$: (i) a Wide-Band Amplifier (WBA, Krohn-Hite 7602M) for U < 400 V and $\tau_r$ > 0.2 $\mu$s; (ii) a high-voltage (HV) amplifier (Trek 2220) for 0.4 kV < U < 2 kV and $\tau_r$ > 50 $\mu$s; (iii) a double-output HV switch (PVM-4210, Directed Energy) for unipolar DC pulses with 0.4 kV < U < 1.9 kV and $\tau_r$ > 0.02 $\mu$s; (iv) finally, a set of home-made transformers adapted to different frequency ranges were used to amplify the WBA output voltage up to about 2 kV for AC bursts with frequency f > 1 kHz. In this way, we could apply fields up to 1 kV/mm to the sample in the whole frequency range of interest. This field limit is imposed not only by the available amplifiers but also by the dielectric breakdown of air in usual laboratory humidity conditions.
However, one feature of our external-electrodes setup allowed us to apply inside the sample, in a transient way, a field twice as large as this limit: Indeed, when a DC voltage U is applied to the external electrodes for a time much longer than the charge relaxation time of the solvent (\(\tau_{ch} \simeq 20\)ms for our sample), the conductivity charges of the solvent move and accumulate on the inner side of the capillary glass wall facing the electrodes. This process proceeds up to the complete screening of the field within the suspension because of the opposite field created by the accumulated charges. If now the voltage applied to the electrodes is rapidly reverted, from U to –U, the external field and the field due to the accumulated charges have the same sign, resulting in a twice stronger transient field in the suspension, 2U/L$_e$. This field relaxes back to zero with the same characteristic time \(\tau_{ch}\) because of the migration of the charges to the opposite wall. Since the rise-time of the TEB signal (\(\tau_{on} \simeq 2 ms\) for the longest stacks) is much smaller than $\tau_{ch}$, we can measure, during the transient regime, the induced birefringence under the internal field $E_\textrm{int}=2U/L_e$ i.e. up to 2 kV/mm. We call this field-inversion procedure the “double-field” trick in the main text.\\	

The field-induced birefringence was measured in real time, under polarizing microscope (Leitz Ortholux II), with the apparatus described in detail in references \cite{Dozov2011b,Buluy2018}. It consists of a stabilized light source, an optical compensator introducing an additional constant phase shift, a photo-multiplier tube (PMT), a load resistor R$_L$ transforming the PMT anode current in a voltage difference, a differential amplifier with band-pass filters (AM 502, Tektronix), and a digital oscilloscope (DSO-X 2004A, Agilent Technologies) that accumulates the signal up to 64000 counts.
Nevertheless, this setup was modified in several ways to achieve the high sensitivity and fast response time required for some of the measurements. For large particle stacks, the signal was strong enough, with good Signal-to-Noise (S/N) ratio, and we used, as previously, a Berek compensator introducing a $\lambda/4$ phase-shift, resulting in a transmitted intensity linearly proportional to the induced birefringence. However, this simple optical configuration is not convenient when the induced birefringence is very small because the PMT current is a sum of the small time-dependent induced-birefringence signal and a large constant term coming from the $\lambda/4$ phase-shift introduced by the compensator. Therefore, the residual noise after removal of the constant term is too high. To remedy this issue, for weak signals, we replaced the Berek compensator with a Senarmont compensator and we uncrossed the analyzer by just a few degrees. In this way, we obtained a more sensitive (quadratic) optical response and a smaller constant term, resulting in significantly better S/N ratio. Moreover, the Senarmont compensator introduces the same phase-shift over the whole field of view of the microscope, allowing us to use a much larger measurement window, which also improves the S/N ratio significantly. 
The response time of the set-up is mainly defined by the R$_L$C$_A$ constant of the PMT anode. For measurements with suspensions of isolated platelets, due to their large rotational diffusion constant, D$^r$ = 6.6$\times$10$^5$s$^{-1}$, the response time must be kept as short as possible. Since the anode capacitance, C$_A\simeq$300 pF, is fixed, we used a load resistor  $R_L\leq 1 k\Omega$. For measurements of stacks (and for the static measurements of isolated particles), we used $R_L = 1 k\Omega$, which affords both a good S/N ratio and an acceptable response time, R$_L$C$_A\simeq$300 ns. For the dynamic experiments with the isolated particles, we improved the time-resolution of the set-up to less than 50 ns by deconvolution of the measured response with the instrumental function measured in a separate experiment. \\

Our external-electrodes technique allowed for long-term, in-situ, time-resolved studies of suspensions of CdSe platelets and their stacks without sample degradation. Indeed, solvent evaporation is impossible in sealed capillaries and the glass wall separating the electrodes from the colloidal dispersion prevents any electrochemical degradation of the sample despite the repeated application of strong fields over several hours. To study the stacking kinetics, we monitored the same capillary over 15 months by measuring its TEB in exactly the same experimental conditions. The first measurement, made at t$_0$ = 0 days after sample preparation and addition of oleic acid, revealed only the presence of isolated platelets (no change was observed in the next experiments for about one week). Later measurements, made at times t$_1$ = 25, t$_2$ = 35, t$_3$ = 62, t$_4$ = 105, and t$_5$ = 415 days, revealed the presence of stacks (labelled respectively St$_i$ for i = 1, 2, \ldots 5) of increasing size. Between the measurements, the capillary was kept horizontal and, due to its small diameter D = 1 mm, no sedimentation was observed, even at time t$_5$. The path length of the probe light in the sample is also 1 mm, which drastically reduces light absorption and scattering from the dichroic platelets. We note that, for a classic Kerr cell, this length is at least ten times larger, so that much smaller concentrations are required, which would drastically lower the stacking rate and therefore make the experiment impossible.\\


We thank Dr Santanu Jana for the synthesis of samples at early stages of the projects and ANR NASTAROD for funding.

\bibliography{paper}

\end{document}